\documentclass[10pt,journal]{IEEEtran}
\usepackage{pdfpages} 
\IEEEoverridecommandlockouts 
\usepackage{blkarray}                                      
\usepackage{algpseudocode}                                 
\usepackage{algorithm}
\renewcommand{\algorithmicrequire}{\textbf{Input:}}        
\renewcommand{\algorithmicensure}{\textbf{Output:}}        
\usepackage{graphicx}                                      
\usepackage{amsmath}
\usepackage{amssymb}
\usepackage{amsfonts}
\usepackage{amsthm}
\usepackage[mathcal]{eucal}
\usepackage{mathrsfs}
\usepackage{booktabs}
\usepackage{enumerate}
\usepackage{multirow}
\usepackage[subrefformat=parens,farskip=0pt,justification=centering]{subfig}
\usepackage{color}
\usepackage{cite}                                          
\usepackage{comment}                                       
\usepackage{soul}                                          
\soulregister\cite7
\soulregister\ref7
\soulregister\pageref7
\usepackage{etoolbox}                                      
\usepackage{url}
\usepackage{nth}                                           
\usepackage{bm}                                            
\usepackage{courier}
\usepackage{balance}
\usepackage{threeparttable}
\usepackage{xcolor,colortbl}
\usepackage{footnote}
\usepackage{listings}
\usepackage{setspace}                                      

\usepackage{verbatim}
\usepackage[bookmarks=false]{hyperref}
\hypersetup{
    colorlinks = true,
    citecolor  = blue,
    linkcolor  = blue,
    urlcolor   = blue,
}
\usepackage{tikz}
\usetikzlibrary{patterns,snakes}
\usetikzlibrary{positioning,calc,fit,decorations.pathmorphing,shapes.geometric, shapes.gates.logic.US, calc}
\usetikzlibrary{arrows,arrows.meta,decorations.markings,shapes,shapes.arrows}
\usetikzlibrary{decorations,decorations.pathreplacing}
\usetikzlibrary{backgrounds}
\usepackage{filecontents}                                  
\usepackage{pgfplots}
\usepackage{pgfplotstable}
\usepackage{scalefnt}
\pgfplotsset{compat=newest}
\usepackage{caption}
\usepackage{cleveref}
\Crefformat{figure}{Fig.~#2#1#3}                           
\Crefname{subfigure}{Fig.}{Figs.}
\Crefname{figure}{Fig.}{Figs.}
\Crefformat{table}{TABLE~#2#1#3}                           
\usepackage[figuresright]{rotating}

\definecolor{CUHKorange}{RGB}{244,106,18} 
\definecolor{CUHKblue}{RGB}{0,111,190}    
\definecolor{CUHKgreen}{RGB}{0,127,128}   
\definecolor{CUHKred}{RGB}{228,46,36}     
\definecolor{CUHKyellow}{RGB}{198,148,34} 
\definecolor{CUHKdark}{RGB}{114,44,114}   
\definecolor{CUHKmiddle}{RGB}{144,44,144} 
\definecolor{CUHKlight}{RGB}{167,44,167} 
\definecolor{CUHKpurple}{RGB}{117,15,109}
\definecolor{CUHKgold}{RGB}{221,163,0}
\definecolor{CUHKribbon}{RGB}{244,223,176}
\definecolor{CUHKblack}{RGB}{34,24,21}

\renewcommand{\bf}[1]{\textbf{#1}}
\renewcommand{\tt}[1]{\texttt{#1}}

\newcommand{\minisection}[1]{\vspace{.1in}\noindent{\textbf{#1}}}

\usepackage{tcolorbox}
\tcbuselibrary{skins,breakable}
    {\endtcolorbox}
    {\endtcolorbox}

\usepackage[top=0.76in,bottom=0.918in,left=0.62in,right=0.62in]{geometry}

\crefname{mytheorem}{Theorem}{Theorems}
\crefname{mylemma}{Lemma}{Lemmas}
\crefname{myclaim}{Claim}{Claims}
\crefname{myproperty}{Property}{Properties}
\crefname{mycorollary}{Corollary}{Corollaries}

\algrenewcommand\textproc{\texttt}

\makeatletter
\let\OldStatex\Statex
\renewcommand{\Statex}[1][3]{%
  \setlength\@tempdima{\algorithmicindent}%
  \OldStatex\hskip\dimexpr#1\@tempdima\relax
}
\makeatother

\RequirePackage[normalem]{ulem} 
\RequirePackage{color}\definecolor{RED}{rgb}{1,0,0}\definecolor{BLUE}{rgb}{0,0,1} 

\begin{document}
\date{}
 


\title{
   Floorplet: Performance-aware Floorplan Framework for Chiplet Integration 
}

\author{
    Shixin Chen,                       \quad
    Shanyi Li,                         \quad
    Zhen Zhuang,                       \quad
    Su Zheng,                          \quad
    Zheng Liang,                        \\
    Tsung-Yi Ho,                       \quad
    Bei Yu,                            \quad
    Alberto L. Sangiovanni-Vincentelli
    \thanks{Shixin Chen, Shanyi Li, Zhen Zhuang, Su Zheng, Tsung-Yi Ho, and Bei Yu are with Chinese University of Hong Kong.}
    \thanks{Zheng Liang and Alberto L.~Sangiovanni-Vincentelli are with University of California, Berkeley.}
}

\maketitle
\pagestyle{plain}

\begin{abstract}

A chiplet is an integrated circuit (IC) that encompasses a well-defined subset of an overall system’s functionality.
In contrast to traditional monolithic system-on-chips (SoCs), chiplet-based architecture can reduce costs and increase reusability, representing a promising avenue for continuing Moore’s Law.
Despite the advantages of multi-chiplet architectures, floorplan design in a chiplet-based architecture has received limited attention.
Conflicts between cost and performance necessitate a trade-off in chiplet floorplan design since additional latency introduced by advanced packaging can decrease performance.
Consequently, balancing performance, cost, area, and reliability is of paramount importance.
To address this challenge, we propose Floorplet (\underline{Floorp}lan chi\underline{plet}), a framework comprising simulation tools for performance reporting and comprehensive models for cost and reliability optimization.
Our framework employs the open-source Gem5 simulator to establish the relationship between performance and floorplan for the first time, guiding the floorplan optimization of multi-chiplet architecture.
The experimental results show that our framework decreases inter-chiplet communication costs by 24.81\%. 

\end{abstract}

\section{Introduction}\label{sec:intro}

\IEEEPARstart{H}{ow}
can we design complex electronic systems with high performance and low cost? 
This is a fundamental question that has driven the development of integrated circuit (IC) design for the past 50 years. 
Moore's Law\cite{moore1998cramming} has been the dominant paradigm that describes how advances in chip technology improve chip performance and reduce cost by integrating more transistors on a single die. 
However, as the cost of finer lithography patterns increases, the economic benefits of Moore's Law are diminishing. 
Therefore, many foundries such as TSMC, Samsung, and Intel are exploring alternative solutions to reduce wafer costs and improve yields~\cite{zhuang2022multi-package}. 
One of the promising solutions is the use of advanced heterogeneous integration and multi-chiplet architecture.

A chiplet is an integrated circuit with a specific function obtained by partitioning a traditional monolithic system-on-chip (SoC). 
Chiplets can also be considered intellectual property (IP) components for reuse in multiple systems to reduce design costs. 
Additionally, chiplets provide an opportunity for heterogeneous integration~\cite{douglas2017heterogeneous, iyer2016heterogeneous, pal2020DSE-chiplet} with different technology nodes. In chiplet-based architecture, less important components can use cheaper technology nodes, while some specified components like analog, power, and memory modules can use more advanced technology to reduce IC costs further and improve yield\cite{sangiovanni2023chiplet-auto-design}.

\begin{figure}[t]
    \hspace{1.6cm}
    \includegraphics[width=0.35
    \textwidth]{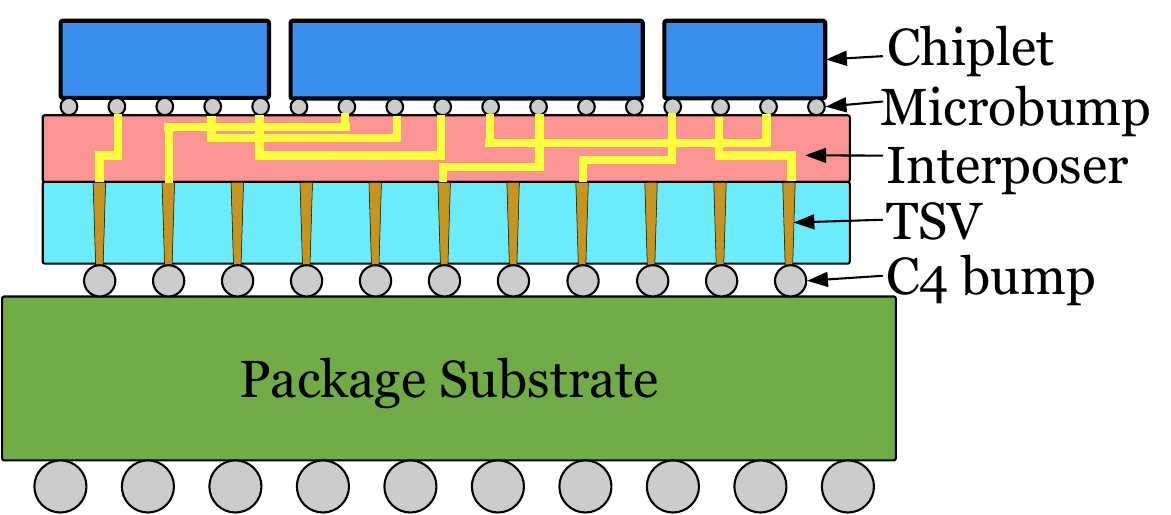}
    \caption{The interposer-based 2.5D package containing chiplets.}
   \label{fig:chiplet-a}
\end{figure}

The chiplet-based design method needs advanced package techniques to fully utilize its characteristics. 
Chiplets can be assembled in three dimensions (3D ICs) or placed on a silicon interposer (2.5D ICs).
\Cref{fig:chiplet-a} illustrates an example of advanced 2.5D package using multi-chiplet architecture\cite{greenhill2017interposer-based}. 
In the figure, chiplets with various functions are placed on a silicon interposer using microbumps.
Data communication between chiplets occurs through wires within the interposer.
The interposer is then bonded to the package substrate using C4 bumps.

\begin{figure}[t]
    \hspace{1.2cm}
    \includegraphics[width=0.30
    \textwidth]{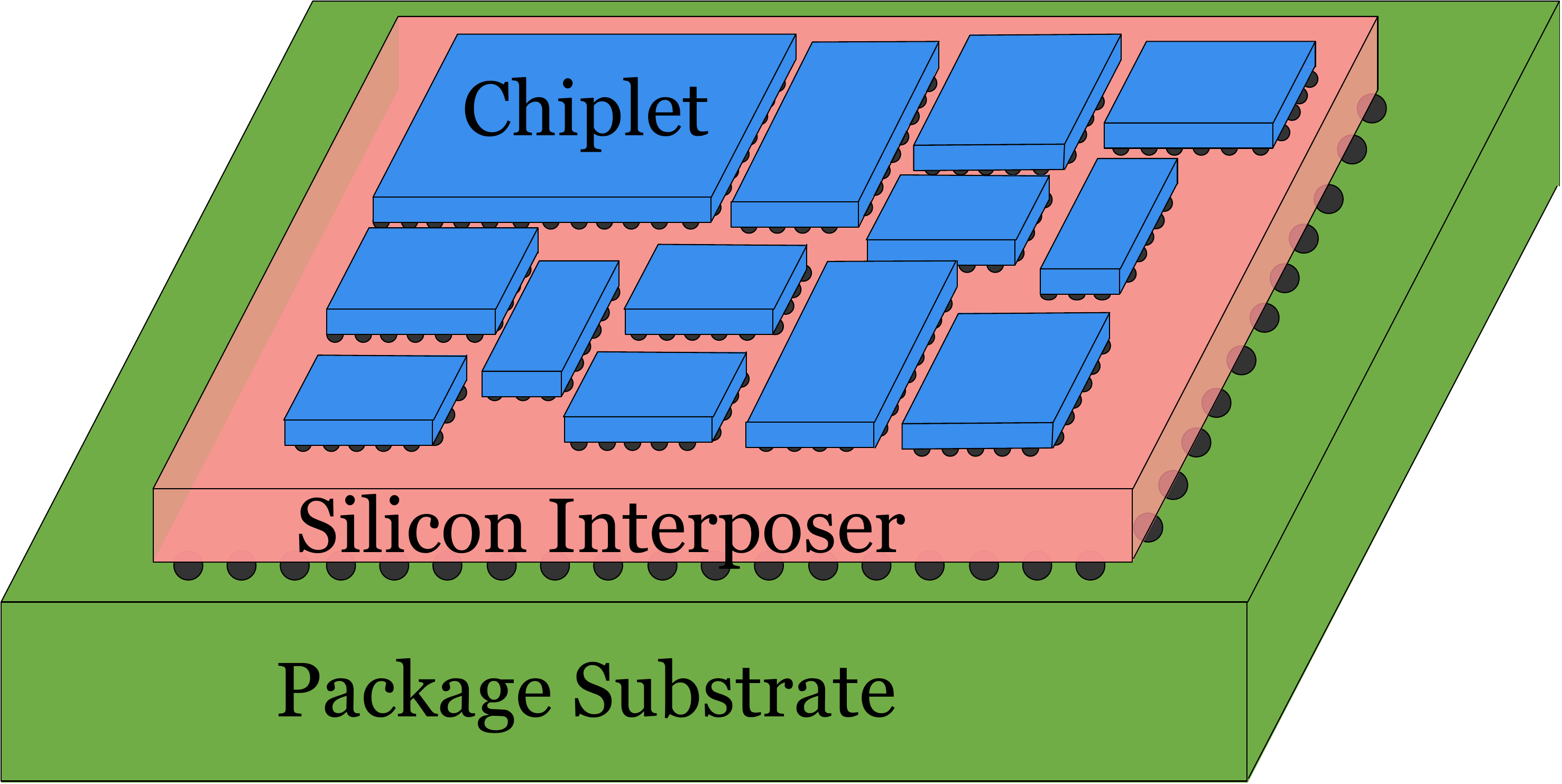}
    \caption{Floorplan design of chiplets in 2.5D package.}
    \label{fig:chiplet-b}
\end{figure}

The multi-chiplet architecture has attracted significant attention from academia and industry due to the advantages mentioned above. 
However, there are also challenges and drawbacks associated with chiplets that hinder their popularity in practical usage. 
If not handled properly, they may result in decreased performance and higher fabrication costs instead of reducing costs.

Firstly, from a performance perspective, the multi-chiplet architecture can suffer from degraded performance due to the extra physical wirelength between chiplets on the interposer. 
Inter-chip nets are routed on the redistribution layers (RDLs) by chip-scale wires\cite{2013interposer}.
Therefore, deciding the proper locations of chiplets on the interposer, \textit{i.e.}, floorplan design of multiple chiplets shown in \Cref{fig:chiplet-b}, will greatly impact the communication between chiplets.
Consequently, the overall performance of chiplet-based architecture is highly dependent on the quality of floorplan design.

Secondly, from a cost standpoint, the use of the advanced package can introduce more reliability issues. 
The reliability issues of the 2.5D package will potentially affect IP functionality and reduce the service life of the chiplet system.
Due to a higher degree of mismatch in the coefficient of thermal expansion (CTE) in advanced packages, there may appear bump reliability affected by stress, and chiplet cracking caused by package warpage\cite{2021chiplet-crack,che2015warpage,chong2021thermal-chiplet}.
 
Thirdly, from design automation tools, we lack design methodologies and electronic design automation (EDA) tools to support practical 2.5 packages specifically. 
Lots of literature on the chiplet-based floorplan design is based on abstract chiplets without specific functionality. 
In \cite{sangiovanni2023chiplet-auto-design}, the automated design of the chiplet was analyzed from a high-level architecture perspective, and the results have not been tested in realistic circuit designs.
Prior art \cite{ahmad2022chiplet-cost-yield} focuses on partitioning SoC into chiplets based on cost analysis. Work \cite{zhuang2022multi-package} simply considers package reliability and ignores performance degradation caused by extra package cost and die-to-die interfaces.
Therefore, designing practical methodologies and EDA tools based on actual circuits is important to boost the performance of chiplet.

To address these challenges and drawbacks, we propose Floorplet, a performance-aware framework for co-optimizing floorplan and performance of chiplet-based architecture.
Unlike previous papers that use abstract chiplets without specific functionality or ignore performance metrics in floorplan design, we consider realistic chiplets with complex data flow and incorporate performance factors into floorplan optimization.
Our framework consists of three main components with corresponding contributions as follows:

\begin{itemize}
    \item \textbf{parChiplet}: An algorithm to partition realistic SoC into functional chiplets that can be fabricated and analyzed by EDA tools. 
    The chiplets have specific functions so that we can bring more hardware information to enable an analysis of complex data flow in chiplet-based architecture. 
    \item \textbf{simChiplet}: A simulation platform based on Gem5 that evaluates the performance impact of different floorplan solutions for chiplet-based architecture. 
    The communication latency of chiplet-to-chiplet depends on the latency in the interposer, which builds a connection between the floorplan design and architecture performance. 
    \item \textbf{optChiplet}: A floorplan framework for chiplet architecture that considers reliability, cost, and area in addition to performance metrics. 
    We tested our framework on various benchmarks to show that it outperforms previous methods and achieves co-optimization of architecture and technology.
\end{itemize}

The remainder of this paper is organized as follows: 
Section \ref{sec:prelim} introduces the cost and reliability model for the chiplet-based architecture and important issues in floorplan design. 
Section \ref{sec:method} provides detailed components of Floorplet and corresponding algorithms.
Section \ref{sec:exp} conducts experiments to demonstrate the superiority of Floorplet and analyzes experimental results. Finally,
Section \ref{sec:conc} concludes the paper.

\section{Preliminaries}\label{sec:prelim}

\subsection{Chiplet and Cost Model}

\minisection{Chiplet Definition}.
A chiplet is a small chip that is obtained by partitioning a large monolithic SoC chip. 
Chiplets can be fabricated using different technologies and integrated into a larger system using advanced package techniques like interposer-based 2.5D packages. 
Chiplets can offer several benefits over monolithic SoC chips, such as lower fabrication and design costs, higher yield, and better performance.
However, chiplet-based architecture also poses several challenges and drawbacks that need to be addressed. 
One of the main challenges is how to optimize the floorplan of chiplets on the interposer, which affects the performance, reliability, cost, area, and power of the system. 

\minisection{Yield and Defect Density}.
According to Moore's Law, the number of transistors in an IC doubles every 18-24 months, which has been well proved in the past half-century \cite{moore1998cramming}. 
However, the size of transistors in IC has now met its physical limit and cannot be reduced as in the past \cite{haensch2006silicon}. 
Therefore, advanced package techniques like interposer-based 2.5D packages are proposed to address this issue to build large-scale systems \cite{feng2022cost}. 
Compared to a large monolithic SoC chip, the 2.5D package contains many smaller chiplets, which are partitioned from the monolithic SoC. 
With chiplets, a larger system can be constructed, and the fabrication and design costs will be reduced substantially.

\begin{figure}[tb!]
  \centering
  \includegraphics[width = 0.94 \linewidth]{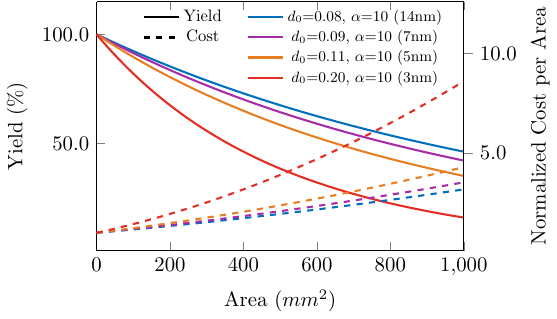}
  \caption{The relation between yield/cost and area of different technologies. 
  The figure shows that the yield decreases and the cost increases as the area increases, especially for advanced technologies with higher defect density.}
  \label{fig:yield-cost}
\end{figure}

One of the main factors that affect the fabrication cost is yield, which is the probability that a chip die is functional and free of defects. 
The yield depends on the die area and the defect density of the technology. 
A widely used model to estimate yield is the negative binomial model \cite{cunningham1990yield-model}:
\begin{equation}
  Y(s)_{chip} = (1+\dfrac{d_{0}s}{\alpha})^{-\alpha},
\end{equation}
where $Y$ is the yield, $s$ is the die area, $d_{0}$ is the defect density and $\alpha$ is the cluster parameter. 
In the 7-nm technology, the typical value of $d_{0}$ is 0.09 $cm^{-2}$, and the typical value of $\alpha$ is 10 \cite{feng2022cost}. 
With this equation, we can estimate the manufacturing costs based on the processed wafer's yield $Y$ and unit price $P_{0}$. 
\Cref{fig:yield-cost} shows that the yield decreases and the cost increases as the area increases, especially for advanced technologies with higher defect density.

We can use Taylor expansion to approximate the manufacturing cost per yielded area as follows: 
\begin{equation}
  y(s)=\dfrac{P_{0}}{Y_{die}} \approx P_{0}(1+d_{0}s+\dfrac{\alpha-1}{2\alpha}d_{0}^{2}s^{2}),
\end{equation}
where $y(s)$ is the manufacturing cost per yielded area, which will rise quickly as the die sizes increase. 
This equation implies that with smaller chiplets partitioned from a monolithic SoC, we can lower the $y(s)$, thereby decreasing overall manufacturing expenses.

\minisection{Cost Model of 2.5D Package}.
In addition to the manufacturing cost of chiplets, we must consider the bonding cost and yield of chiplets and the interposer in the 2.5D package.
As shown in \Cref{fig:chiplet-b}, the raw cost of a 2.5D package consists of several components: raw chiplets $C_{chip}$, raw substrate $C_{sub}$, raw package $C_{pack}$, raw silicon interposer $C_{inter}$ and bonding cost $C_{bond}$ of each chiplet. 
The bonding process involves attaching chiplets to an interposer and the interposer to a substrate using micro-bumps or through-silicon vias (TSVs). 
The bonding process may introduce defects or failures that affect the functionality of the package.

Therefore, by considering raw cost and bonding cost, the overall cost of the 2.5D package can be expressed as follows:  

\begin{equation}
  \begin{aligned}
C_{package}&=C_{pack}\\
       &+C_{inter}\times (\dfrac{1}{y_{1} \times y_{2}^{n}\times y_{3}}-1)\\
       &+ C_{sub} \times(\dfrac{1}{y_3}-1),
  \end{aligned} 
\end{equation}
where $n$ is the number of chiplets, $y_{1}$ is the yield of the interposer, $y_{2}$ is the bonding yield of chiplets, $y_{3}$ is the bonding yield of interposer.

The overall cost of the 2.5D package is 
\begin{equation}
   C_{2.5D}=\dfrac{\frac{C_{pack}}{Y_{pack}}+\sum_{i \in n}(\frac{C_{chip\_i}}{Y_{chip\_i}}+C_{bond\_i} )} {Y_{bond}^{n}},
   \label{equ:package_cost}
\end{equation}
where $Y_{pack}, Y_{chip}, Y_{bond}$ are respectively the yield of the package, the yield of chiplets, and the yield of the bond. 
The values of each term in our cost model are validated based on data from public information and in-house sources \cite{feng2022cost}. 

\subsection{Reliability Issues of 2.5D Package}

\minisection{Warpage}.
Warpage is a major reliability concern in advanced packages that refers to abnormal bending of shape due to the mismatch in the CTE of different materials.
Warpage can cause the package to bend and result in the cracking of chiplets and substrates.
The critical parameters affecting warpage include mold thickness, molding materials, and the ratio of chiplet to package area \cite{2021chiplet-crack,che2015warpage,chong2021thermal-chiplet}.
Warpage can be measured by the variation in vertical height at different positions on a package. 
In this study, we aim to control warpage from the center to the edges of packages using an effective warpage computing model introduced in previous works \cite{irwin2021warpage-model, tsai2019theoretical}. 
The warpage model is given by:

\begin{equation}
  w(x)=\dfrac{t \cdot \Delta \alpha \cdot \Delta T}{2 \cdot \lambda \cdot D}[\frac{1}{2} x^{2}-\dfrac{cosh(kx)-1}{k^{2}cosh(kl)}].
  \label{equ:warpage}
\end{equation}
where $\Delta\alpha$ represents the difference in CTE between the chiplets and the substrate, $\Delta T$ represents the thermal load.
The coefficients $t, \lambda, D$, and $k$ are related to the material properties. 
The center of the package is used as the origin for building this model. 
The variables $x$ and $l$ represent half the length of a chiplet and half the length of a substrate, respectively.

\minisection{Bump Stress}.
Bump stress is another major reliability concern in advanced packages that refers to the mechanical stress induced by bumps during the bonding process. 
Bump stress can increase the risk of bump cracks or component deformation, which can affect the functionality of the package. 
Bump stress depends on the location and size of bumps, as well as the placement and shape of components near the bumps.

Hotspot bumps are defined as the bumps at the edge of the silicon interposer that have a higher risk of deformation than the bumps in the interposer center with uniform stress. 
Analysis has shown that the edges of bumps experience high stress due to the proximity of components, especially for hotspot bumps that have a higher possibility of deformation or crack \cite{jung2012bump-stress,sakuma20213bump-issue}. 
In this work, we aim to control the bump stress by avoiding the overlap between chiplets and hotspot bumps. 
We use a margin space around hotspot bumps to prevent chiplets from overlapping with them and reduce bump stress.

\subsection{Floorplan of Chiplets}\label{sec:prelim-floorplan}

\begin{figure*}[t]
\begin{center}
\begin{minipage}[b]{0.35\textwidth}
  \includegraphics[width = 1 \textwidth]{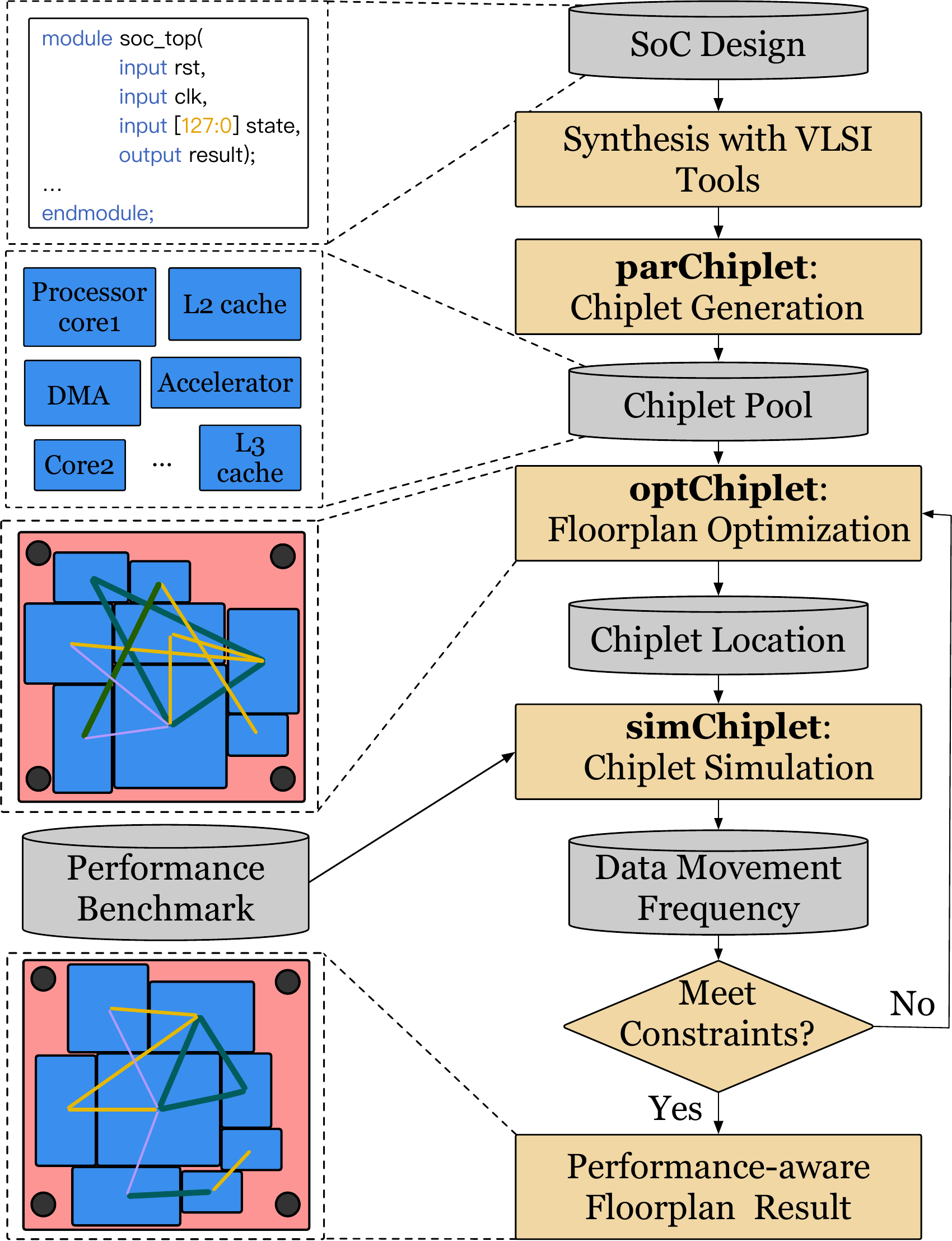}
  \caption{The overall flow of the Floorplet.}
  \label{fig:overall_flow}
\end{minipage}\hspace{7pt}
\begin{minipage}[b]{0.63\textwidth}
  \includegraphics[width = 1 \textwidth]{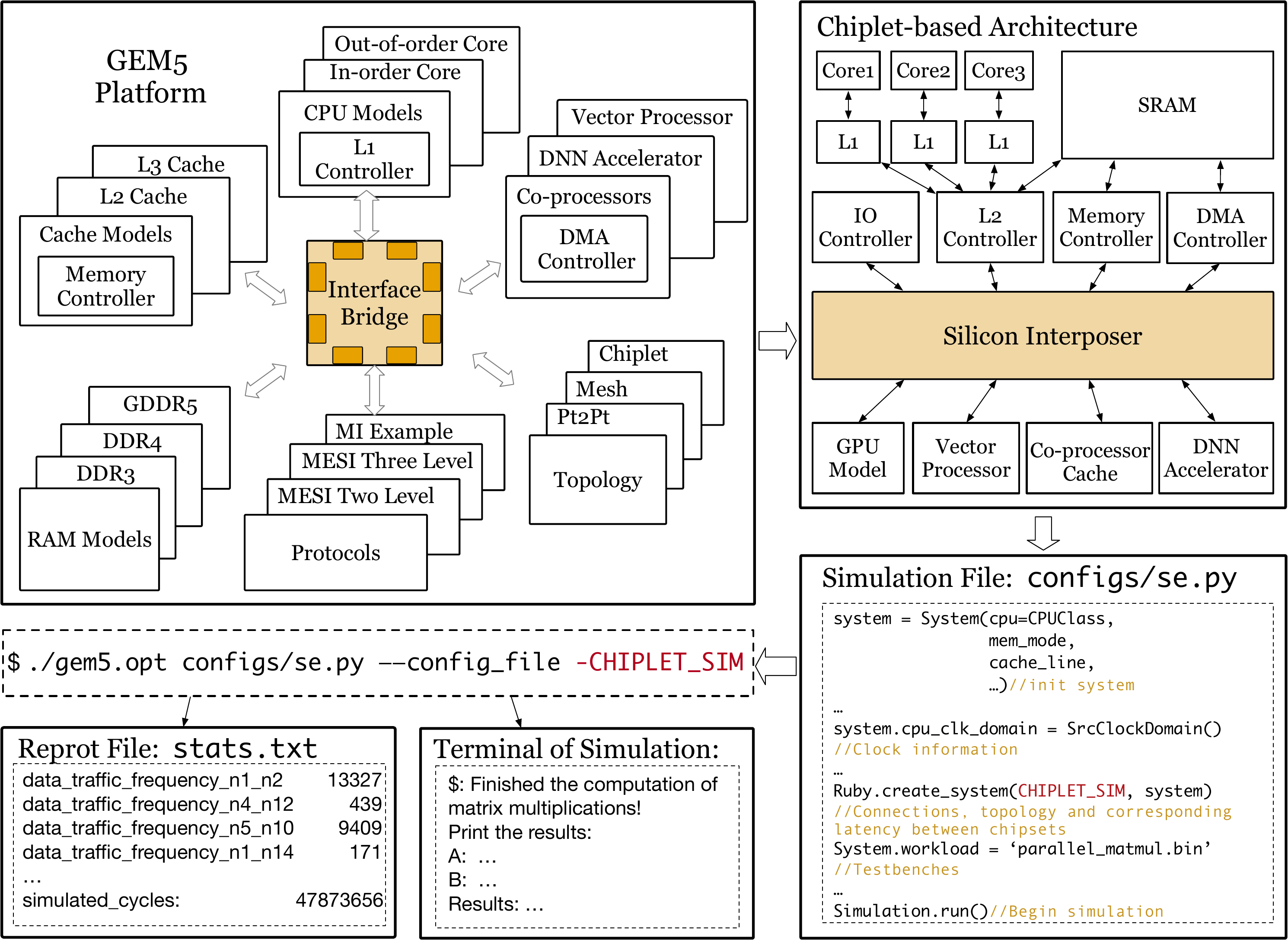}
  \caption{The simulation flow embedded to Gem5 platform.}
  \label{fig:simulation_flow}
\end{minipage}

\end{center}
\end{figure*}

The floorplan is an important stage in designing chiplet-based systems that allocates the major functional blocks on the interposer. 
The floorplan affects the overall area, the routing wirelength, the thermal effects, and the reliability of the design. 
Prior works~\cite{liu2014chiplet-floorplan, osmolovskyi2018chiplet-placement} only consider the area and wirelength of the floorplan, while ignoring the performance metrics, such as the data movement frequency and communication latency between different chiplets. 
If the optimization of the floorplan can take the performance metrics into consideration, the final performance of the chiplet system can be improved.

Work \cite{Chiplet-GIA-ICCAD2022-Li} utilizes the modular design scheme, which means that each chiplet is placed on the tiles with an equal size on the interposer. 
The scheme improves the re-usability of each chiplet IP, while it may increase the potential of extensive interposer area cost due to some tiles being free on the interposer.
In our framework, we adopt a different scheme for chiplet floorplanning.
The floorplan scheme in our work does not involve routers in the interposer, instead, we attempt to add more flexibility to the routing wire in the interposer and minimize the interposer area.
In this way, the chiplet-based architecture can benefit from the router-free scheme with a larger floorplan solution space to obtain the optimal floorplan solution.

Chiplet-based architecture makes it possible to reduce time to market by utilizing existing IPs. 
By combining various chiplets with different functions in the 2.5D package, design companies can create new electronic systems without starting from designing and testing to manufacturing SoC. 
For example, in the recent AMD ZEN 2 architecture~\cite{naffziger20202amd-chiplet}, server and desktop processors can share
the identical chiplet named core complex die (CCD), which contains the CPU cores and caches. 
However, given such existing chiplets, it is challenging to achieve a trade-off between cost and performance. 
Previous works mainly focus on the chiplet cost model, but the performance model lacks exploration. 
In our framework, we combine the performance metric with the floorplan of the chiplet. 
We build a chiplet simulation platform to evaluate the chiplet design and elaborate it in \Cref{sec:method}.

\subsection{Problem Formulation}

The input of our framework is a set of chiplets $\mathcal{C}$ partitioned from a monolithic SoC, a netlist $N$ that connects the chiplets, and the bump positions of the silicon interposer. 
The output of our framework is a floorplanning solution that provides the location and orientation of each chiplet on the silicon interposer. 
The objective of the optimization is:
\begin{equation}
  \begin{aligned}
  &\text{min }\beta_{1} \times wl + \beta_{2} \times \alpha_{p} + \beta_{3} \times C_{2.5D}, \\
  &\text{s.t. } wpg_{p} \leq wpgt_{p},
  \label{equ:objective}
\end{aligned}
\end{equation}
where $wl$ represents half-perimeter wirelength (HPWL) between chiplets, 

$\alpha_{p}$ represents the warpage of the package, and $C_{2.5D}$ is the cost of 2.5D package from \Cref{equ:package_cost}.
$\beta_{i}$ represents a user-defined coefficient that can be modified for a tradeoff between multiple objectives.

During optimization, three types of constraints are considered:
\begin{itemize}
  \item Overlap constraints: Each chiplet cannot overlap with other chiplets or hotspot bumps.
  \item Warpage constraints: As shown in Equation \eqref{equ:objective}, 
  the warpage of the package $p$ cannot exceed a threshold in both the x-axis and y-axis directions.
  \item Bump margin constraints: A margin space is defined around hotspot bumps to avoid overlap with chiplets and reduce bump stress.
\end{itemize}

\section{Floorplet Framework}\label{sec:method}

\subsection{Overview of the Framework}\label{sec:method-overview}

\Cref{fig:overall_flow} shows the overview of the proposed Floorplet, which consists of three main steps: chiplet partitioning, floorplan optimization, and performance evaluation. 

Firstly, we propose the \textbf{parChiplet} algorithm to partition a monolithic SoC into a set of chiplets with different functions and area constraints. The algorithm recursively divides the hierarchical tree structure that represents the SoC components and their connections.

Secondly, we propose \textbf{optChiplet} to formulate the floorplan of the chiplet as a mathematical programming (MP) problem that considers multiple objectives such as package cost, wirelength, warpage, and bump stress. 
We solve the MP problem to obtain a primary floorplan solution that provides the location and orientation of each chiplet on the silicon interposer.

Thirdly, we evaluate the performance of the chiplet design using \textbf{simChiplet} built based on Gem5 garnet3.0\cite{Bharadwaj2020heterogarnet}. The simulation platform can report the data movement frequency and latency of different chiplet pairs based on various benchmarks. 
We use the data movement frequency as feedback to further optimize the floorplan results. 
In the simulation platform, the data are sent or received in packages, which have an equal volume. In this way, more data movement frequency can be seen as more data movement volume.
The details of each step will be elaborated as follows.

\subsection{\textbf{parChiplet}: Chiplet Generation Method}\label{sec:method-parChiplet}

\begin{algorithm}[t]
    \small
	\renewcommand{\algorithmicrequire}{\textbf{Input:}}
	\renewcommand{\algorithmicensure}{\textbf{Output:}}
	\caption{\texttt{parChiplet($T,a_{\min}, a_{\max}$)}}
	\label{alg1}
	\begin{algorithmic}[1]
	    \Require ($T, a_{\min}, a_{\max}$), $T$ is the hierarchical tree structure like \Cref{fig:soc-tree}, $a_{\min}$ is the minimal threshold of chiplet area, $a_{\max}$ is the maximal threshold of chiplet area. 
	    \Ensure The chipet pool $\mathcal{C}$.
        \State $\mathcal{C}=\emptyset$;
        \State $N=child(T)$;\Comment{\footnotesize $child(\cdot)$ gives the children nodes of $T$. }
        \State $n_{r}=\emptyset$
        \For {$n_{i}\in N$}
        \If{$area(n_{i})>a_{\max}$}\Comment{\footnotesize $area(\cdot)$ gives node's area.}
        \State \texttt{parChiplet($n_{i}, a_{\min}, a_{\max}$)};
        \ElsIf { $area(n_{i})<a_{\min}$}
        \State $n_{r}=comb(n_{i},n_{r})$; \Comment{\footnotesize $comb(\cdot)$ combines two nodes.}
        \Else
        \State $\mathcal{C}=\mathcal{C} \bigcup \{n_{i}\}$;
        \EndIf\\
        $\mathcal{C}=\mathcal{C} \bigcup \{n_{r}\}$;
        \EndFor 
        \\
        \Return $\mathcal{C}$;
	\end{algorithmic}
 \label{alg:parChiplet}
\end{algorithm}

Unlike previous works \cite{zhuang2022multi-package}\cite{sangiovanni2023chiplet-auto-design} that target on highly abstract chiplets without specific functions, we choose an actual hardware design of SoC and utilize synthesis tools in very-large-scale-integration (VLSI) flow to obtain the area and netlist of the design. Then, based on various functions and area information, we design an algorithm to partition chiplets from the SoC and obtain a chiplet pool. 

\begin{figure}[tb!]
    \centering
    \includegraphics[width=0.98\linewidth]{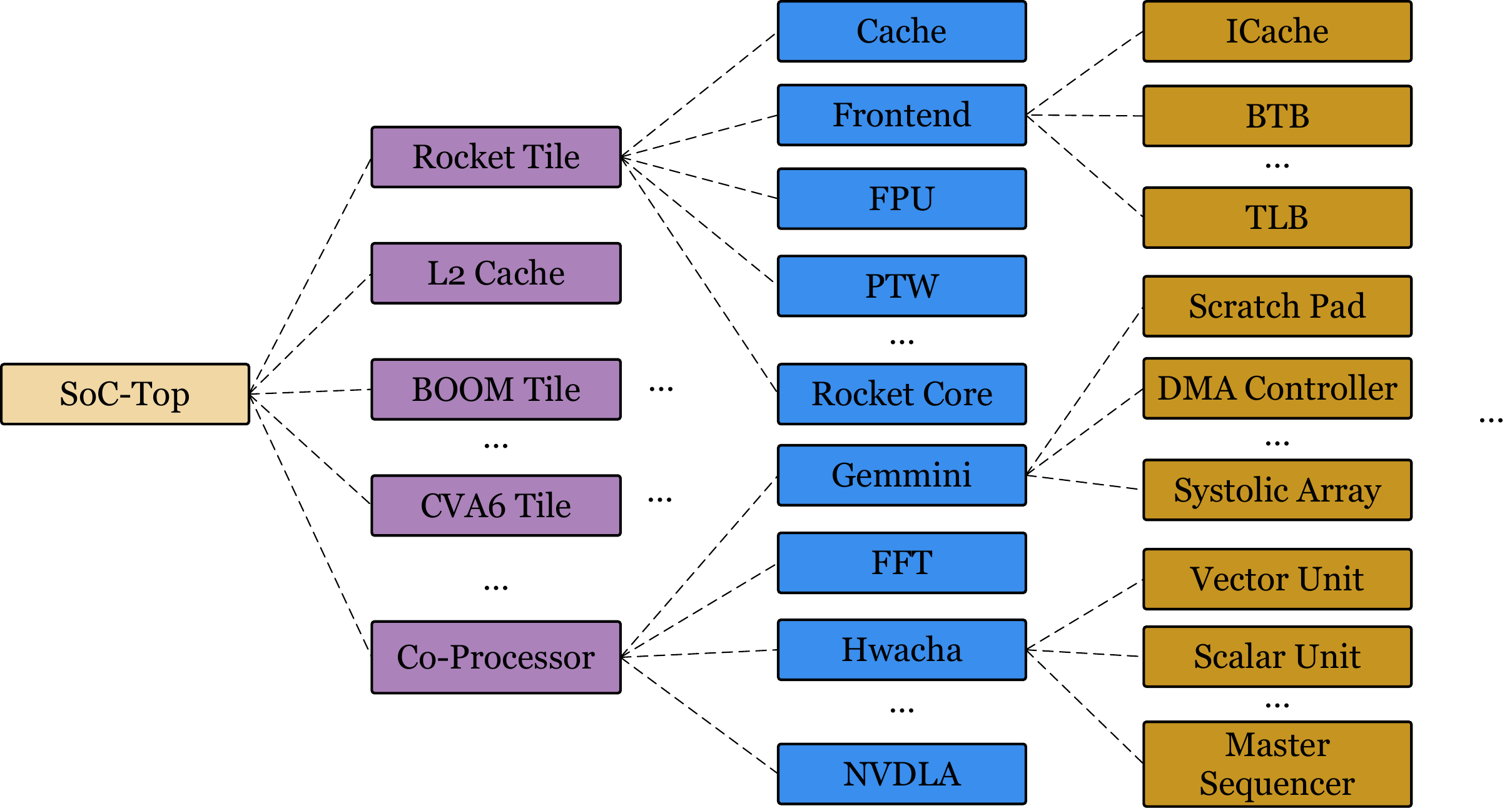}
    \caption{The hierarchical tree of SoC components.}
    \label{fig:soc-tree}
\end{figure}
Some previous works like~\cite{zhuang2022multi-package} ignored the integrity of chiplet functions when obtaining chiplets, \textit{i.e.}, some circuit macros that work tightly may be partitioned to distinctive chiplets. 
To bring more practical design information into chiplet design, we use Python language to design a script to process the synthesis result of SoC from very-large-scale integration (VLSI) tool like Hammer\cite{liew2022hammer}.
The output of the script is a hierarchical tree of SoC, and an example is shown in \Cref{fig:soc-tree}.
The hierarchical tree takes functional integrity and area into consideration. 
In this way, the partitioned chiplets have relatively independent functions that can be reused as IPs by other designs. 
Meanwhile, we can control the number of chiplets divided from the SoC system within an acceptable range.  

Furthermore, the granularity of the partition is very important during constructing a chiplet pool.
On the one hand, if the chiplet area is too large like a part of a monolithic SoC, the benefit from the decreasing of area cost will be eliminated. 
On the other hand, if the chiplet area is too small like a fine-grained fragmentation with incomplete function, it is impossible for fabrication to provide such an IP.

Therefore, we design the rule to make sure the area range$(a_{\min},a_{\max})$ is given as:
\begin{equation}
    a_{\min}=\dfrac{a_{SoC}\times rr}{|\mathcal{C}|_{\max}}
    ,\quad
    a_{\max}=\dfrac{a_{SoC}\times rr}{|\mathcal{C}|_{\min}}
    ,
\end{equation}
where $|\mathcal{C}|_{\min},|\mathcal{C}|_{\max}$ represent the typical minimal and maximal number of chiplets in chiplet-based architecture, respectively. 
$a_{SoC}, rr$ represent the overall area of the monolithic SoC given by the synthesis tool and the relaxation ratio of $a_{SoC}$, respectively. $rr$ is set to make sure to avoid the failure of the chiplet partition.
The total area of partitioned chiplets should be larger than the original $a_{SoC}$ to avoid failure of the floorplan because extra areas of interfaces die-to-die (D2D) are introduced.

By combining all analysis into an algorithm, we propose \Cref{alg:parChiplet}.
The algorithm is straightforward to process the SoC hierarchical tree recurrently.
The input of the algorithm is the synthesis results given by Hammer\cite{liew2022hammer} from different SoC designs. 
For various SoC designs, once given the synthesis results, the hierarchical tree can be constructed automatically with function integrity and area information.
Since we have designed the parser to analyze the chiplet interconnection relationship, the algorithm can be generalized across different SoC.

\subsection{\textbf{simChiplet}: Chiplet Simulation Method}\label{sec:method-simChiplet}

\begin{figure}[t]
    	\subfloat[]{\includegraphics[width=.52\columnwidth]{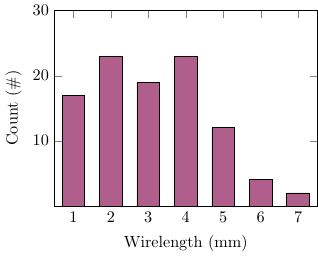}
        \label{fig:latecny-wire-a}}\hspace{-4pt} 
    	\subfloat[]{\includegraphics[width=.45\columnwidth]{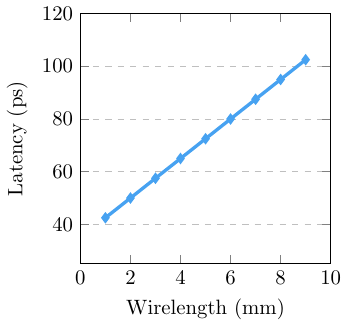}
        \label{fig:latecny-wire-b}}\\
    	\caption{The latency-wirelength model.
     (a) The distribution of the wirelength between chiplets.
     (b) Latency weight \textit{v.s.} wirelength relation
    	}
            \label{fig:latecny-wire}        
      \end{figure}

In this step, we build a platform based on Gem5 garnet3.0\cite{Bharadwaj2020heterogarnet} to evaluate the data movement frequency of the chiplet-based architecture and the overall latency of running various workloads. 
The detailed simulation flow is shown in \Cref{fig:simulation_flow}, where we set the characteristics of various hardware components to mimic the functionality of the original monolithic SoC.
In the original simulation tool, network-on-chip (NoC) in Gem5 garnet3.0 combines different modules in the chiplet-based architecture.
The NoC module in gem5 is used to record the data request and data volume, which can assist in obtaining the data movement between chiplets.
However, the original framework does not support setting various latencies between different chiplet-pair flexibly, so we improve the flexibility of the original topology by embedding the latency weight given by latency-wirelength model in \Cref{fig:latecny-wire-b}, which is going to be introduced as follows. 

In the 2.5D package, the inter-chiplet communication latency is determined by the length of the routed wires inside the interposer, the microbumps, and the die-to-die interfaces.
In the monolithic SoC, the communication latency is determined by the critical path in the circuit with a typical max wirelength of about 1.4 mm. 
However, in the chiplet-based architecture, the max wirelength can be much longer than that of SoC. 
For example, in the chiplet-based architecture containing 64 cores \cite{kim2020chiplet-64core-riscv}, the max wirelength can reach about 10 mm.

We cannot ignore the latency introduced by the chiplet architecture, as it can affect the performance of the system. 
In UCIe 1.0 \cite{sharma2022UCIe}, an open industry standard for on-package connectivity between chiplets, the latency of the interface should be smaller than 2 ns. 
If the system runs with 2GHz, extra clock cycles will be introduced to chiplet communication.
To show the latency influence, we choose an SoC containing 2 Rockets cores and 2 BOOM cores and use their HPWL distribution to estimate their latency weights as shown in \Cref{fig:latecny-wire}.

Work \cite{kim2020chiplet-64core-riscv} designed and constructed a chiplet-based 64-core processor to illustrate the chiplet design flow.
It built and verified the interposer delay model consisting of resistance, inductance, conductance, and capacitance (RICC).  
According to the results, in the 0.2 to 10.0mm length range, as the wire becomes longer, both communication delay and energy increase linearly. 
In our experiment, the wire length falls into the wire length range mentioned above, so we think this conclusion can be utilized in our framework.
Since the work has verified their conclusion, we suggest using this conclusion to build our interposer delay model.

Therefore, according to the HPWL between the chiplets on the silicon interposer and the relationship between the latency and wirelength\cite{kim2020chiplet-64core-riscv}, we map the wirelength into six ranges in \Cref{fig:latecny-wire-a}, and each range has a specific latency weight.
With the latency-wirelength model in \Cref{fig:latecny-wire-b},   
we can embed a file containing latency information and connections of chiplet to our simulation platform with an option \texttt{--CHIPLET\_SIM} in the command line shown in \Cref{fig:simulation_flow}. 

\subsection{\textbf{optChiplet}: Floorplan Optimization Method}\label{sec:method-optChiplet}
The goal of our floorplan framework is to optimize the placement of chiplets on the silicon interposer while minimizing wirelength and improving reliability. 
In this section, we present our mathematical programming (MP) models for solving the floorplan problem. We first introduce the primary floorplan model that considers the chiplet dimensions, locations, rotations, warpage, and bump stress.
Then we describe the performance-aware floorplan model that incorporates the data movement frequency between chiplets obtained from our simulation platform.

\begin{figure}[t]
	\centering
    	\subfloat[Initialization]{
            \includegraphics[width=.15\textwidth]{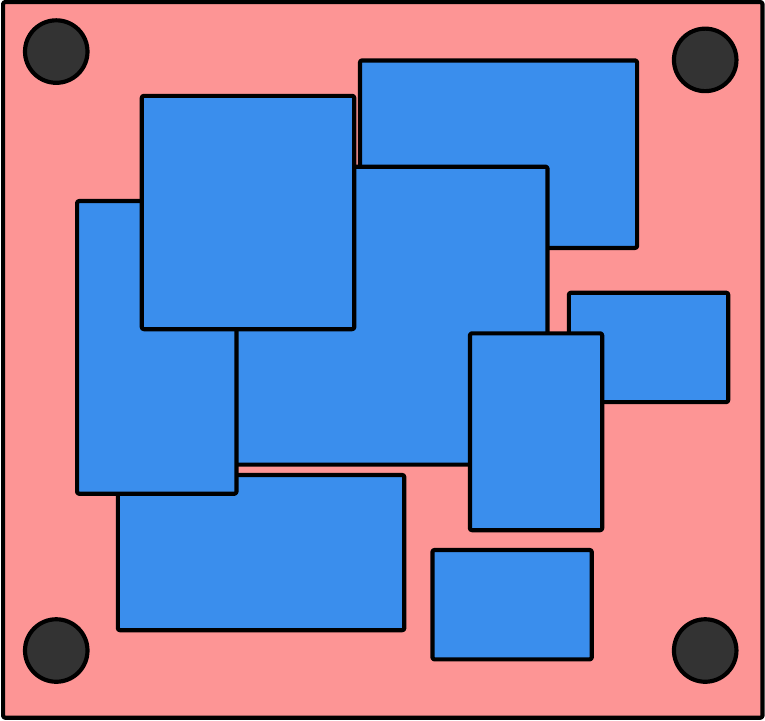}   \label{fig:floorplan-a}
        } 
    	\subfloat[Before]{
            \includegraphics[width=.15\textwidth]{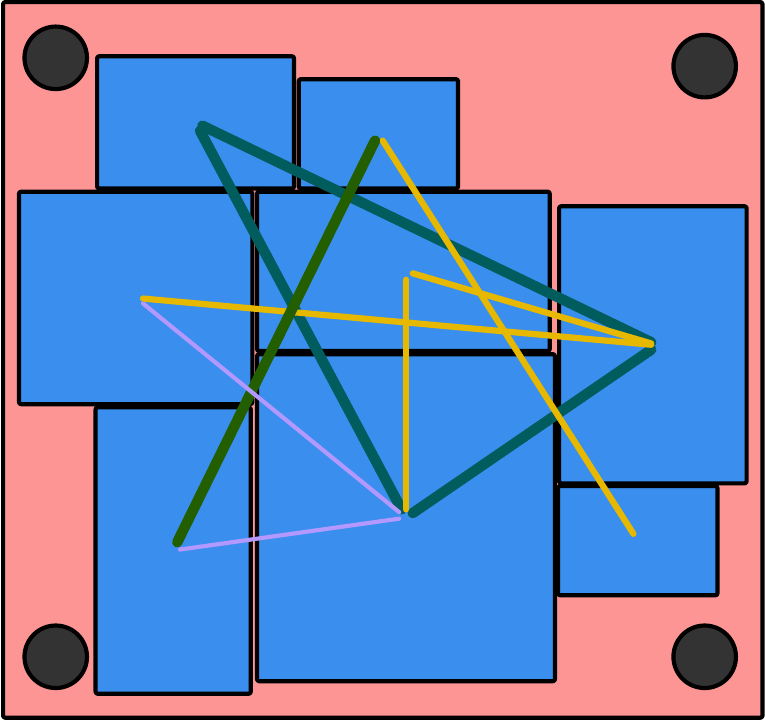} \label{fig:floorplan-b}
        }
    	\subfloat[After]{
            \includegraphics[width=.15\textwidth]{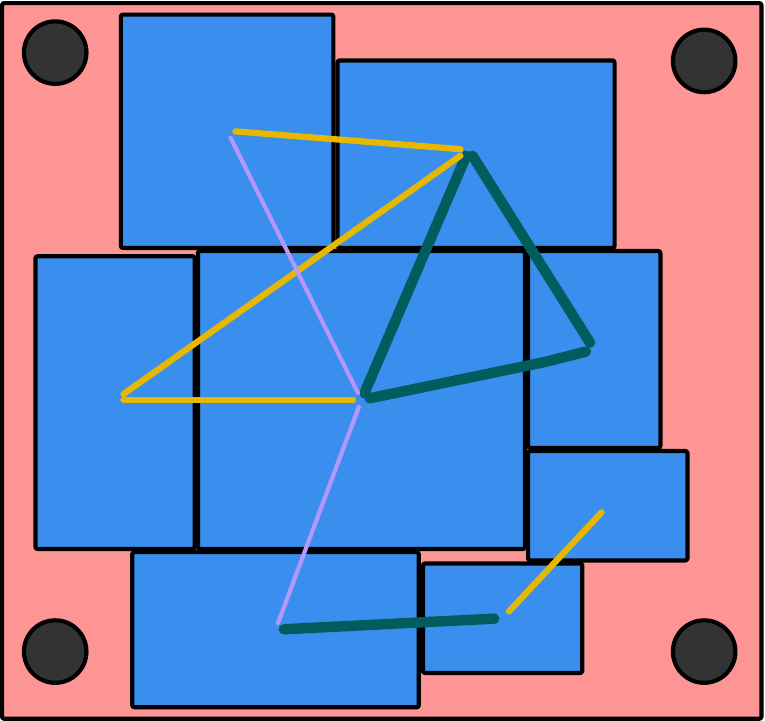}  \label{fig:floorplan-c}
        }
    	\caption{The different floorplan designs of chiplet-based architecture. (a) The macros, bumps, nets, and fixed-outline boards are generated from input data. (b) The floorplan design with eight chiplets without performance metrics. The wider line represents higher data movement frequency between chiplets. (c) The performance-aware floorplan design. }
            \label{fig:floorplan}        
\end{figure}

\minisection{Primary floorplan}.
We assume that the input data consists of $n$ chiplets $\mathcal{C}=\{c_{1},c_{2},\cdots,c_{n}\}$, each with a fixed outline and a variable location and orientation. The center area of the silicon interposer is a bounding box $W \times H$, where we aim to place the chiplets without overlap. We use $x_{c_{k}}$ and $y_{c_{k}}$ to denote the x-coordinate and y-coordinate of the lower-left corner of chiplet $c_{k}$, respectively. We also use $w_{c_{k}}$ and $h_{c_{k}}$ to denote the width and height of chiplet $c_{k}$, which depend on whether it is rotated or not.

To formulate the floorplan problem as an MP model, we introduce some auxiliary variables and constraints as follows:

\begin{table}[t!]
    \centering
    \caption{The notations used in\textbf{optChiplet} }
    \label{tab:notations}
     \resizebox{0.99\linewidth}{!}
     {  
     
        \begin{tabular}{c|c}
          \toprule
           {Notations}  & {Meaning } \\
            \midrule
            $\mathcal{C}$                      & all chiplets of the architecture \\
            $c_{k} \in \mathcal{C}$            & a chiplet $k$ in the architecture\\
            $p$                                & the package of the design\\
            $W,H$                              & width and height of floorplan region\\
            $(x_{c_{k}},y_{c_{k}})$            & coordinate of the lower-left corner of chiplet $c_{k}$\\
            $(x_{b_{s}},y_{b_{s}}$             & coordinate of the center of bump $b_{s}$\\
           $w_{c_{k},h_{c_{k}}}$               & width and height of chiplet $c_{k}$\\
            $(p_{i,j},q_{i,j})$                & indication of relative position of $c_{i}$ and $c_{j}$\\
            $r_{k}$                            & indication of rotation of $c_{k}$\\
            $d_{cc}$                           & radius of the circumscribed circle of chiplet\\
            $d_{mr}$                           & radius of the margin region around each bump\\
            $wpg_{p}^{x},wpg_{p}^{y}$          & warpage of package $p$ in $x$-axis and $y$-axis\\

            \bottomrule
        \end{tabular}
        
    }
\end{table}
\begin{itemize}

  \item 
In practical chiplet-based architecture, there exists a small distance to allow routing wire between two near chiplets.
After the floorplan stages, there are follow-up stages like routing stages for the final chiplet-based design fabrication. 
Therefore, we simplified the optimization process by omitting the small distance between two nearby chiplet.
To prevent overlap between chiplets, we use binary variables $p_{i,j}$ and $q_{i,j}$ to indicate the relative positions of chiplet $c_{i}$ and chiplet $c_{j}$. The non-overlap constraints can be expressed as:
\begin{align}
    \label{equ: no-overlap-1}
    x_{c_{i}}+w_{c_{i}}&\leq x_{c_{j}} + W \cdot (p_{i,j}+q_{i,j}),\\
    y_{c_{i}}+h_{c_{i}}&\leq y_{c_{j}} + H\cdot(1+p_{i,j}-q_{i,j}),\\
    x_{c_{i}}-w_{c_{i}}&\geq x_{c_{j}}-W\cdot (1-p_{i,j}+q_{i,j}),\\
    y_{c_{i}}-h_{c_{i}}&\geq y_{c_{j}}-H\cdot(2-p_{i,j}-q_{i,j}),\\
    \label{equ: no-overlap-2}
    p_{i,j},q_{i,j}&\in\{0,1\},  1\leq i < j \leq |\mathcal{C}|.
\end{align}
If $(p_{i,j},q_{i,j})=(0,0)$, the chiplet $c_{i}$ is constrained to place on the left of chiplet $c_{j}$; 
if $(p_{i,j},q_{i,j})=(0,1)$, the chiplet $c_{i}$ is constrained to place on the bottom of chiplet $c_{j}$; 
if $(p_{i,j},q_{i,j})=(1,0)$, the chiplet $c_{i}$ is constrained to place on the bottom of chiplet $c_{j}$; 
if $(p_{i,j},q_{i,j})=(1,1)$, the chiplet $c_{i}$ is constrained to place on the top of chiplet $c_{j}$.
In other words, by optimizing this pair, we can change the location of various chiplet to avoid overlap between chiplets.

  \item To allow rotation of chiplets, we use binary variables $r_{{k}}$ to indicate whether chiplet $c_{{k}}$ is rotated by 90 degrees or not. The width and height of chiplet $c_{{k}}$ can be calculated as:
\begin{align}
    w_{{k}} = r_{{k}}\cdot h_{{k}}^{o}+(1-r_{{k}})\cdot w_{{k}}^{o},\\
    h_{{k}} = r_{{k}}\cdot w_{{k}}^{o}+(1-r_{{k}})\cdot h_{{k}}^{o},
\end{align}
where $w_{{k}}^{o}$ and $h_{{k}}^{o}$ are the original width and height of chiplet $c_{{k}}$, respectively.

\end{itemize}

\begin{figure}[t]
\centering
    \includegraphics[width = 0.43\textwidth]{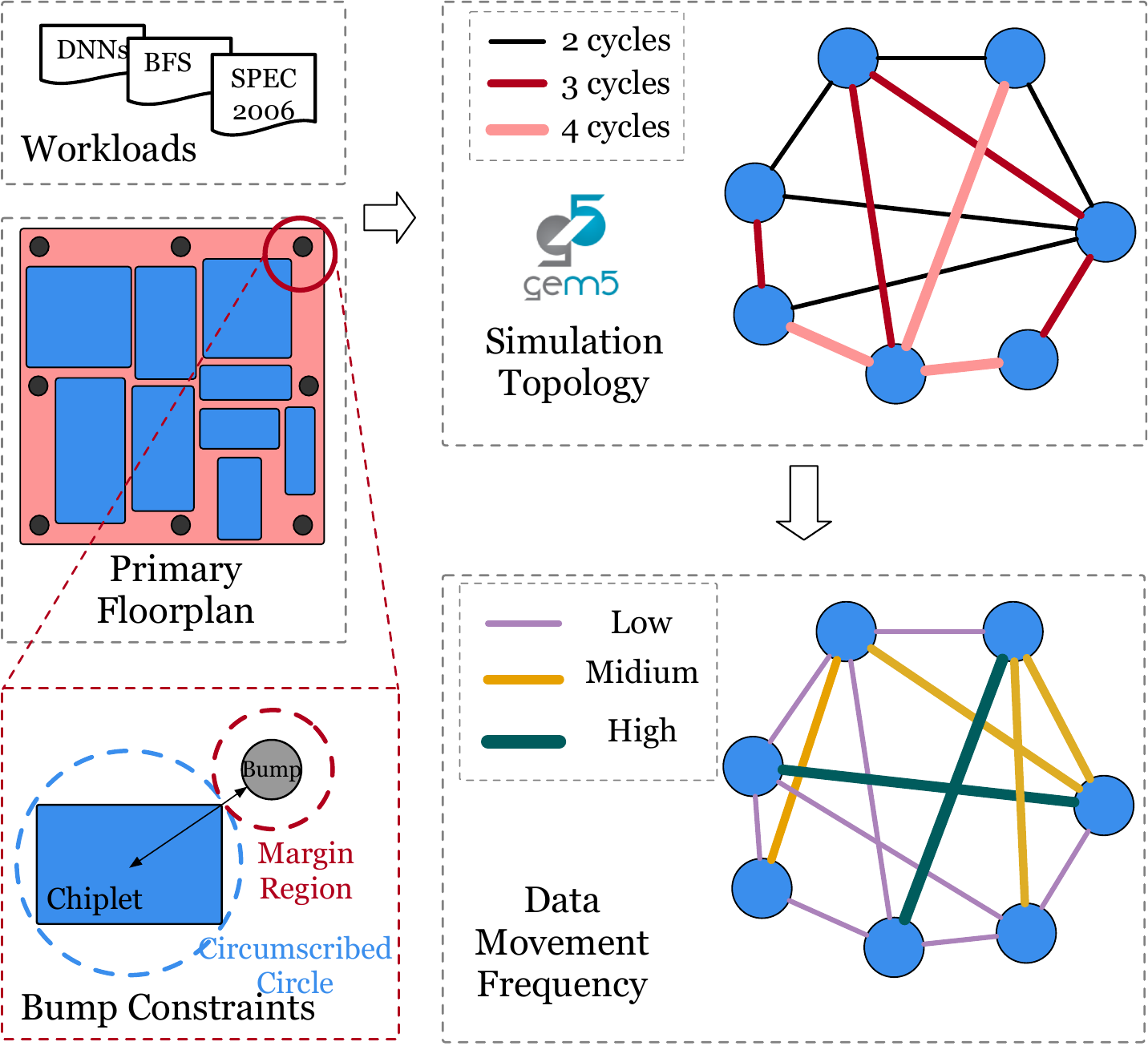}
    \caption{The simulation framework for chiplet-based architecture. The blue circles represent chiplets, while the lines between them represent latency. After the simulation, the data movement frequency between chiplets will be reported, where the wider line represents a higher data movement frequency.}
    \label{fig:chiplet-simulation}
\end{figure}

To account for reliability issues caused by warpage and bump stress, we use continuous variables $wpg_{p}^x$ and $wpg_{p}^y$ to represent the warpage on the x-axis and y-axis directions of the whole package $p$, respectively. The bump constraints are shown in the lower-left corner of \Cref{fig:chiplet-simulation}, where we use continuous variables $d_{cc}$ and $d_{mr}$ to represent the radius of the circumscribed circle of each chiplet and the radius of the margin region around each hotspot bump, respectively. The warpage constraints can be expressed as follows:
\begin{equation}
       w(x)=\dfrac{t \cdot \Delta \alpha \cdot \Delta T}{2 \cdot \lambda \cdot D}[\frac{1}{2} x^{2}-\dfrac{((kx)^{2}+1)/2-1}{k^{2}cosh(kl)}],
\end{equation}
where $t$, $\Delta \alpha$, $\Delta T$, $\lambda$, and $D$ are physical parameters related to the packing materials and dimensions. The warpage in each direction can be calculated by plugging in the corresponding values of $x$. The warpage upper bound can be enforced as:
\begin{equation}
    wpg_p^x\le wpgt_p ,\quad wpg_p^y\le wpgt_p, 
\end{equation}
where $wpgt_p$ is a user-defined threshold for acceptable warpage. The bump margin constraints can be expressed as follows:
\begin{equation}
   (x_{c_{i}}-x_{b_{s}})^2+(y_{c_{i}}-y_{b_{s}})^2 \geq (d_{cc}+d_{mr})^2,
   \label{equ: bump constraints}
\end{equation}
where $(x_{b_{s}},y_{b_{s}})$ are the coordinates of the center of hotspot bump $b_{s}$, and $(x_{c_{i}},y_{c_{i}})$ are the coordinates of the center of chiplet $c_{i}$. This constraint ensures that there is enough spacing between each chiplet and each hotspot bump to avoid excessive stress.

The objective function of the primary floorplan model is to minimize a weighted sum of wirelength ($wl$), area ($\alpha_{p}$), warpage ($wpg_{p}^x$ and $wpg_{p}^y$), and cost of 2.5D package ($C_{2.5D}$) with bump constraints, which can be expressed as:
\begin{equation}
    \beta_{1} wl + \beta_2 \alpha_p + \beta_{3}(wpg_{p}^x + wpg_{p}^y) + \beta_4 C_{2.5D},
\end{equation}
where $\beta_1$, $\beta_2$, $\beta_3$, and $\beta_4$ are user-defined coefficients that reflect different design priorities.

We solve this MP model using an off-the-shelf solver to obtain an initial floorplan solution that satisfies all the constraints and optimizes all the objectives.
\Cref{fig:floorplan-a} shows a set of chiplets partitioned from a monolithic SoC with their connections represented by lines with different widths indicating their data movement frequency.
\Cref{fig:floorplan-b} shows an example of a primary floorplan solution with 8 chiplets placed on a silicon interposer.

\minisection{Performance-aware floorplan}.
After obtaining the primary floorplan solution, we feed it into our simulation platform to evaluate its performance in terms of data movement frequency between chiplets, as shown in \Cref{fig:chiplet-simulation}.

The simulation platform models the application workload, communication patterns, and memory hierarchy of the chiplet-based architecture. The simulation will report a set of frequency values for each pair of chiplets, denoted by $F=\{f_1,f_2,\cdots,f_M\}, M=C_2^{|\mathcal{C}|}$ represents two-combination of $\mathcal{C}$.

We use these frequency values as inputs for our performance-aware floorplan model, which aims to further optimize the placement of chiplets by reducing the latency between frequently communicating pairs. The performance-aware floorplan model has the same variables and constraints as the primary floorplan model, except for an additional term in the objective function that reflects the data movement frequency:
\begin{equation}
    \beta_{1}wl + \beta_{2}\alpha_{p} + \beta_{3}(wpg_{p}^x + wpg_{p}^y) + \beta_{4} +\gamma_{1}\sum_{i}f_{i} 
    \label{equ:final-objective}
\end{equation}
where $\gamma_{1}$ is a user-defined coefficient that controls the trade-off between data movement frequency and other objectives.

We also solve this MP model using an existing solver to obtain a final floorplan solution that balances multiple objectives and meets the performance constraints.
\Cref{fig:floorplan-c} shows an example of a performance-aware floorplan solution with 8 chiplets placed on a silicon interposer.

\section{Experiments \& Analysis}\label{sec:exp}

In this section, we evaluate the effectiveness of our proposed floorplan framework on realistic chiplet-based architectures. We conduct our framework with actual chiplets considering performance or reliability issues, unlike prior works that use abstract rectangles to represent chiplets. We also analyze the tradeoff between different objectives and do some ablation studies to check the effectiveness of reliability constraints.

\subsection{Benchmarks and Baseline}\label{sec:exp-benchmark}

We take advantage of Chipyard\cite{amid2020chipyard}, an open-source SoC generator framework, to generate various SoCs consisting of different hardware components, such as CPU cores (\textit{e.g.}, Rocket\cite{asanovic2016rocket}, BOOM\cite{asanovic2015BOOM}), co-processors (\textit{e.g.}, Gemmini\cite{genc2021gemmini}, Hwacha\cite{lee2015hwacha}), \textit{etc.}
To obtain the area information of each chiplet, we utilize Hammer\cite{liew2022hammer} tools with 7-nm standard cell library ASAP7\cite{vashishtha2017asap7} to synthesize various SoC designs. 
The area information of each chiplet provides the input of  \textbf{parChiplet} to guide the floorplan optimization.

Each SoC can be partitioned into about 10 $\sim$ 30 chiplets based on their functions and build a chiplet pool consisting of about 300 chiplets. 

The chiplet pool is constructed with various SoCs to avoid time-consuming synthesis design. 
Once we meet the same chiplet components, \textit{i.e.,} IPs, in the new chiplet-based architecture, we can reuse the physical information.
Building a chiplet pool provides an opportunity to reuse the synthesis results from Hammer to avoid the extra process of synthesizing. 

For practical chiplet fabrication, it is very time-consuming to decide on the specific design for each chiplet. 
Only after we obtain the final design of the chiplet, the width, and height can be fixed.
For our experiment, we can not carefully design each chiplet one by one.
Therefore, we simplified the process by generating chiplets with a random width/height ratio within a rational range.
Our framework is capable of handling practical chiplet whatever the width/height ratio.
Some selected chiplets are listed in \Cref{tab:chiplets}.

\begin{table}[tb!]
    \centering
    \caption{The selected chiplets from the chiplet pool constructed from various SoC designs}
    \label{tab:chiplets}
     \resizebox{0.82\linewidth}{!}
     {
        \begin{tabular}{c|c|c}
          \toprule
            {Modules} & {Chiplets}  & {Area ($\mu m^{2}$)} \\
            \midrule
            \multirow{6}{*}{ Cores }
            &SmallRocket                      & 561513\\
            &MediumRocket                     & 635068\\
            &LargeRocket                      & 878987\\
            &SmallBOOM                        & 1067891\\
            &MediumRocket                     & 2660359\\
            &LargeBOOM                        & 4083778\\
            \midrule
            \multirow{5}{*}{Co-processors}
            &Systolic Array             & 1330437\\
            &Gemmini Accelerator1       & 2162646\\
            &Gemmini Accelerator2       & 4354390 \\
            &Hwacha  Accelerator        & 1335859\\
            &FFT Accelerator                         & 26445\\
            \midrule
            \multirow{5}{*}{Caches}
            &DMA Controller                  & 212861 \\ 
            &L2 Cache Bank\_1                & 417095 \\
            &L2 Cache Bank\_2                & 557209 \\
            &L3 Cache Bank\_1                & 835814 \\ 
            &L3 Cache Bank\_2                & 3608173 \\
            \bottomrule
        \end{tabular}
        
    }
\end{table}

To boost the computation capability of the SoC architecture, different SoCs are designed specifically with the target of optimizing different applications. 
For instance, \cite{SoC-ISCA2021-Jang} specifically designs the architecture for optimizing DNN applications on flagship mobile devices. 
\cite{SoC-JSSC2011-Sridhara} designs an SoC architecture for mapping medical applications.
Therefore, to optimize the performance of these function-specific chiplet-based architectures, we design the corresponding workloads to do evaluations.

For chiplet-based DNN accelerators, we utilize ResNet-15\cite{he2016resnet}, and MobileNet-v2\cite{howard2017mobilenets} as the workloads. We also use the parallel matrix multiplication workloads, which are the basic operations of DNN-targeting SoC. 
For chiplet-based general-purpose processors, we utilize some representative workloads from the commonly used CPU benchmark SPEC2006\cite{henning2006spec}.
The benchmarks encompass a diverse set of applications with varying performance characteristics, effectively covering the stats of all chiplets.
The workloads for various chiplet-based architectures are listed in \Cref{tab:workloads}.

\begin{table}[tb!]
    \centering
    \caption{The workloads for different chiplet-based architecture from RISC-V benchmark, SPEC2006 and DNNs orderly.}
    \label{tab:workloads}
    \renewcommand{\arraystretch}{1.0}
    \resizebox{.99\linewidth}{!}
    {
        
        \begin{tabular}{c|c|c}
            \toprule
            Architecture & Benchmark & Application\\
            \midrule
            \multirow{ 5}{*}{ $\mathcal{C}_{8}$} 
             & whetstone   &  computer benchmarks\cite{SoC-Whetstone-CJ1976} \\
             & fir2sim     &  DSP-oriented algorithms\\
             & iir         &  DSP-oriented algorithms\\
             & mt-vvad     &  ISA basic instructions\\
             & add-int     &  ISA basic instructions\\
            \midrule
            \multirow{ 4}{*}{ $\mathcal{C}_{16},\mathcal{C}_{22}$ } 
             & 400.perlbench &  email tools in Perl\\
             & 401.bzip2     &  file compression \\
             & 403.gcc       &  C language compiler\\
             & 445.gobmk     &  go game \\
            \midrule
            \multirow{ 3}{*}{ $\mathcal{C}_{30}$ } & ResNet15 &  DNN workload \cite{he2016resnet}\\
            & MobileNet-v2   &  DNN workload\cite{howard2017mobilenets}\\
            & Encoder Module &  DNN modules\cite{vaswani2017transformer}   \\ 
            \bottomrule
        \end{tabular}
        
    }
\end{table}


We compare our framework with the MP-based solver method \cite{zhuang2022multi-package}, which can give the primary floorplan solution without performance consideration. 
We also do some ablation studies to evaluate the effectiveness of warpage constraints and bump stress constraints. 

We implement our framework and the baseline methods in C++ and use Gurobi \cite{2018gurobi} as the MP solver. All experiments are conducted on a Linux machine with an Intel(R) Xeon(R) CPU (E5-2630 v2@2.60GH) and 256 GB RAM.

\subsection{Experiment Setting \& Results}
\label{sec:exp-result}
\begin{table*}[t!]
    \centering
    \caption{The comparison of the methods w./w.o. performance metric }
    \label{tab:results}
    \resizebox{0.99\linewidth}{!}
    {
        \begin{tabular}{c|c|c|c|c|c|c}
            \toprule
            {Architecture} & {Testcases} & {HPWL ($\mu m$)} & {PA ($\times 10^{5}\mu m^{2}$)} & {WPG ($\times 10^{-3} \mu m$)} & {ComCost ($\times 10^{5} $)} & { Average latency($\times 10^{9}$ cycles)}\\ 
            \midrule
            \multirow{4}{*}{ ICCAD'22~\cite{zhuang2022multi-package}}
            &$\mathcal{C}_{8}$            & 19743      & 1.114   & 0.116   & 1.269  & 6.05\\
            &$\mathcal{C}_{16}$           & 107935     & 18.13   & 0.2898  & 4.565  & 81.99   \\
            &$\mathcal{C}_{22}$           & 376158     & 64.25   & 2.996   & 14.476 & 100.5\\
            &$\mathcal{C}_{30}$           & 552483     & 115.1   & 7.065   & 37.37  & 738.9\\
            \midrule
            \multirow{4}{*}{Ours}
            &$\mathcal{C}_{8}$            & 19918      & 1.126    & 0.1189   & 1.068  & 5.60\\   
            &$\mathcal{C}_{16}$           & 106021     & 18.14    & 0.2947   & 2.882  & 73.49\\
            &$\mathcal{C}_{22}$           & 375430     & 65.44    & 3.019    & 12.10  & 83.6\\
            &$\mathcal{C}_{30}$           & 545819     & 115.48   & 6.997    & 26.53  & 605.1 \\
            \midrule
            \multirow{4}{*}{Ratio}
            &$\mathcal{C}_{8}$            & 0.89\%     & 1.14\%    & 2.5\%    & -17.5\%   & -7.4\%\\
            &$\mathcal{C}_{16}$           & -1.77\%    & 0.1\%     & 1.69\%   & -36.8\%   & -10.3\%\\
            &$\mathcal{C}_{22}$           & -0.19\%    & 1.85\%    & 0.76\%   & -16.4\%   & -16.8\%\\
            &$\mathcal{C}_{30}$           & -1.21\%    & 0.35\%    & -1.25\%  & -28.4\%   & -18.1\%\\
            \midrule
            Average
            &N/A                         & \textbf{-0.57\%}    & 0.86\%  & 0.93\%  & \textbf{-24.81\%}  & \textbf{-13.18\%} \\
            \bottomrule
        \end{tabular}
    }
\end{table*}

\begin{table}[t!]
    \centering
    \caption{Ablation study of reliability constraints}
     \label{tab:reliability}
     \resizebox{0.99\linewidth}{!}
    {
        \begin{tabular}{c|c|c}
          \toprule
 { Bump Constraints} & {Normalized Bump Stress}   & {HPWL ($\mu m$)}  \\             
          \midrule
          \multirow{1}{*}{Non-control}
          &4.985                  & 241016\\
         \multirow{1}{*}{Control}
          &0.926                  & 257571\\
           \midrule
          \multirow{1}{*}{Ratio}
          &\textbf{-81.42\%}               & 6.87\% \\
         \bottomrule
         \toprule
         { Warpage Constraints} & {WPG ($\times 10^{-3}\mu m$)}   & {PA ($\times 10^{5}\mu m^{2}$)} \\
         \midrule
         \multirow{1}{*}{Non-control}
          &1.139                  & 31.723\\ 
         \multirow{1}{*}{Control}
          &1.115                   & 31.613\\
           \midrule
         \multirow{1}{*}{Ratio}
          &\textbf{-2.1\%}                 & -0.347\%\\ 
          \bottomrule
        \end{tabular}
        }
\end{table}

We test our framework on four SoC architectures with different configurations, which are partitioned into 8, 16, 22 and 30 chiplets represented by $\mathcal{C}_{8}, \mathcal{C}_{16}$, $\mathcal{C}_{22}$ and $\mathcal{C}_{30}$. 
The $\mathcal{C}_{8},\mathcal{C}_{16},\mathcal{C}_{22},\mathcal{C}_{30}$ are partitioned from a SmallRocket core with the corresponding caches and peripheral modules, a MediumBOOM core with the corresponding caches and peripheral modules, a LargeBOOM core with peripheral modules, and a Gemmini accelerator with the processor core, caches and corresponding DMA controllers.

The data movement frequency between chiplets is the average of various benchmarks that are suitable for the specific SoC design. 
The user-define coefficients are set based on some pre-experiments to obtain a good tradeoff between multiple objectives and ensure convergence of optimization.
According to pre-experiments, $\beta_{1}$, $\beta_{2}$, $\beta_{3}$, $\beta_4$, and $\gamma_{1}$ in \Cref{equ:objective} are set to 1, 10, 100, 1, and 1, respectively. 
The floorplan results demonstrate this setting can achieve a balance between multiple objectives.

In \Cref{tab:results}, we list the experimental results, where HPWL, PA, WPG, ComCost, and Latency represent HPWL wirelength for chiplet routing, package area, warpage of package and inter-chiplet communication cost, and the average clock cycles of finishing the workloads, respectively.
The communication cost is the multiplication of the data movement frequency with the wirelength between two chiplets.
Longer wirelength will bring more latency in RDLs in chiplet-based architecture.
If frequent data movement occurs between two chiplet with too long distances, the overall performance of the design will deduct heavily.
Because data movement costs can bring more waiting time between components.
Therefore, by using this metric, we can evaluate that our method reduces the communication cost.

Our framework can reduce communication costs by placing chiplets with high data movement frequency close to each other. 
As a result, the performance-aware floorplan can decrease the average clock cycles of finishing workloads by 13.18\%. 
The column Latency in \Cref{tab:results} shows that the performance improvement is more significant for architectures with more chiplets.

\begin{figure}[t!]
    \centering
    \subfloat[The primary floorplan solution ($|\mathcal{C}|$=16).]{
        \includegraphics[width=.38\columnwidth]{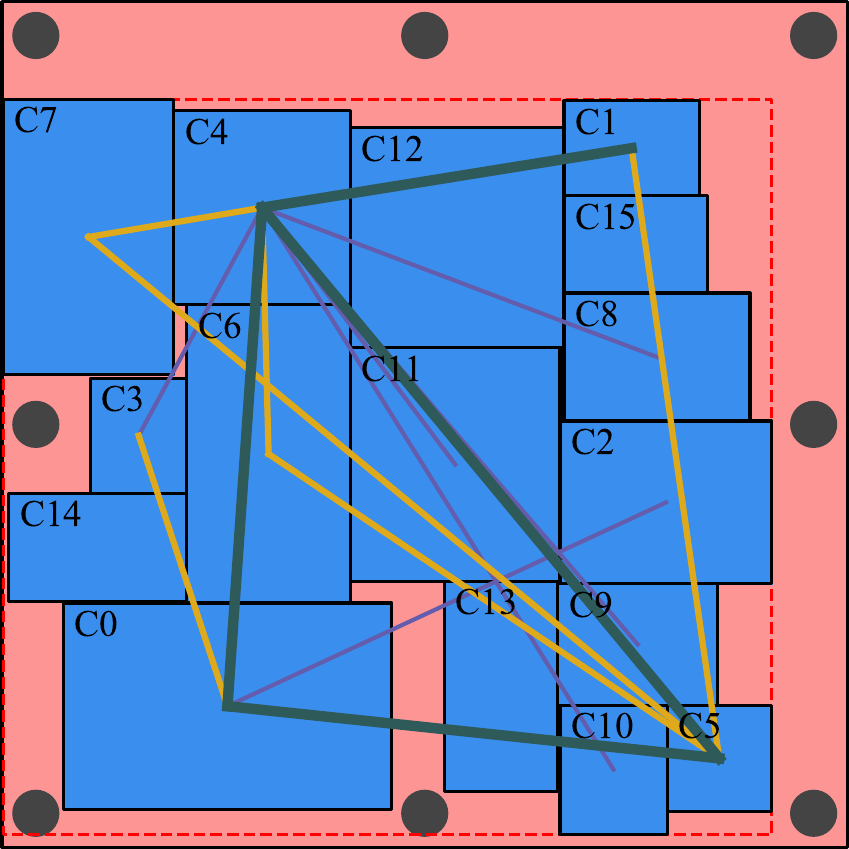}     \label{fig:floorplan_result-a}}
    \hspace{4pt}
    \subfloat[The performance-aware floorplan solution ($|\mathcal{C}|$=16).]{
        \includegraphics[width=.38\columnwidth]{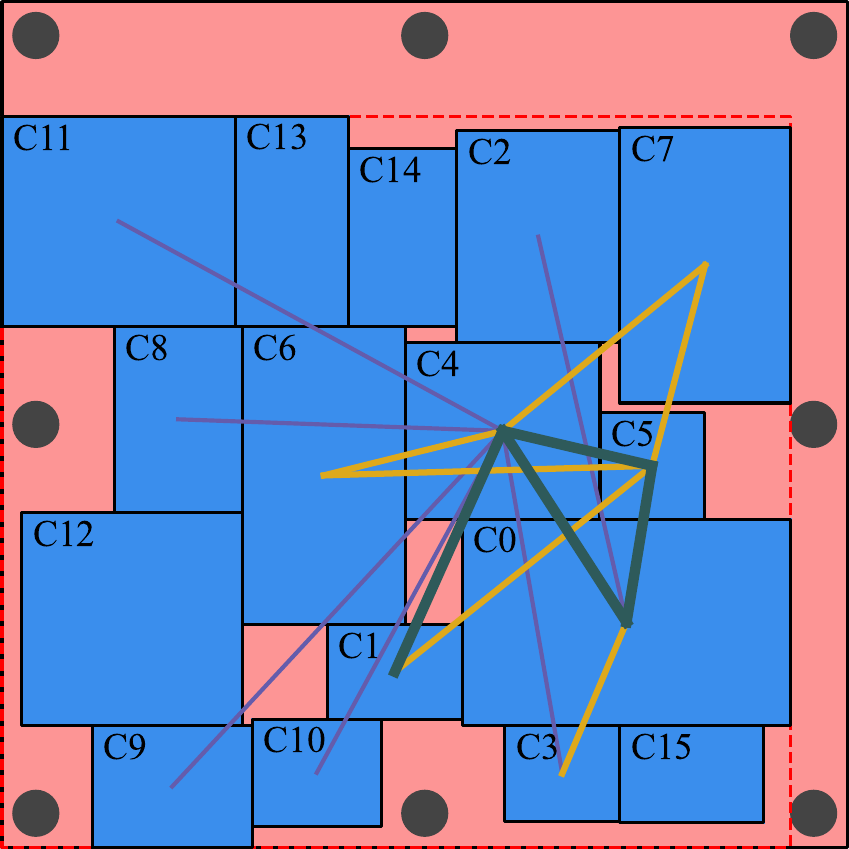}      \label{fig:floorplan_result-b}
    }\\
    \subfloat[The primary floorplan solution ($|\mathcal{C}|$=22).]{
        \includegraphics[width=.38\columnwidth]{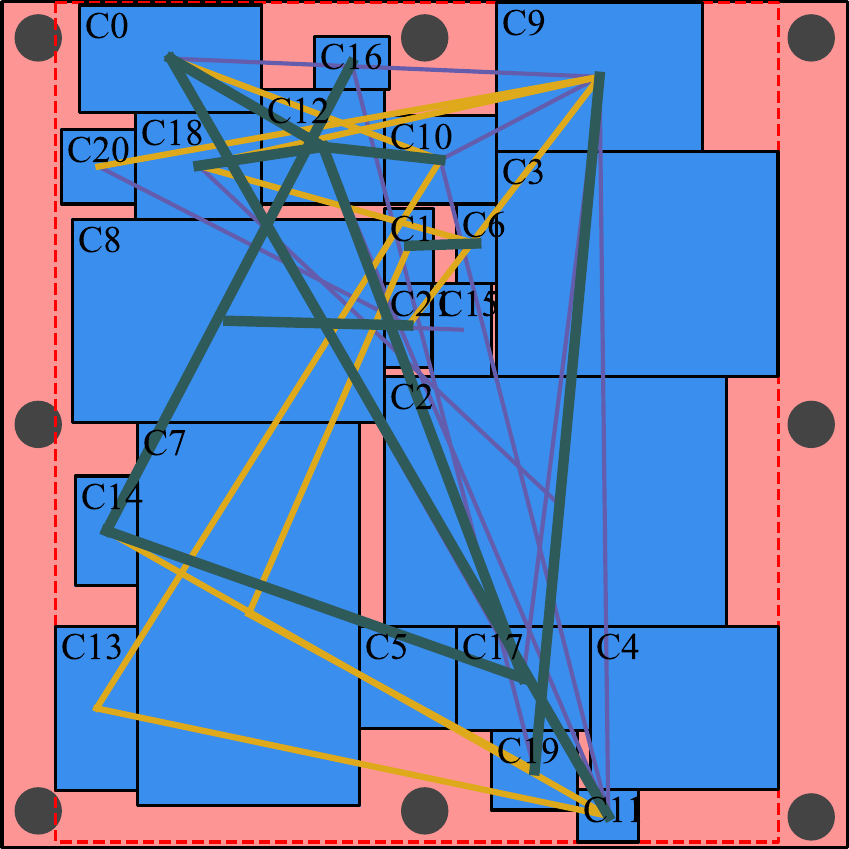}     \label{fig:floorplan_result-c}
    }\hspace{4pt}
    \subfloat[The performance-aware floorplan solution ($|\mathcal{C}|$=22).]{
        \includegraphics[width=.38\columnwidth]{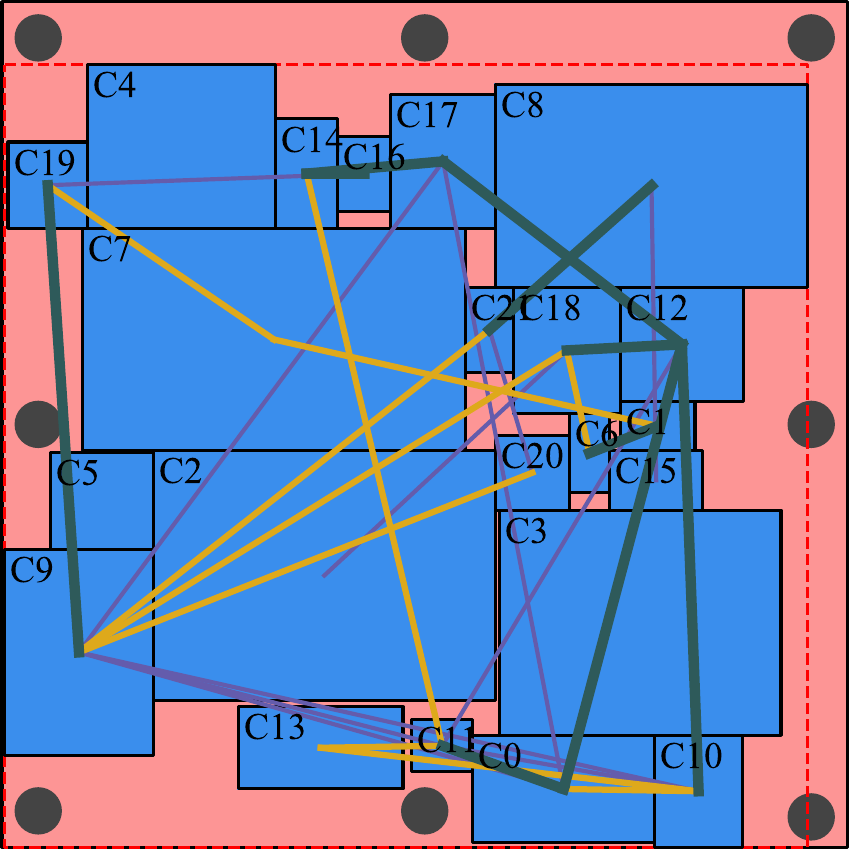}      \label{fig:floorplan_result-d}
    }
    \caption{
        The floorplan design of the chiplet-based architecture ($\mathcal{C}$=16) and ($\mathcal{C}$=22). Our framework decreases communication costs by 36.8\% and 16.41\%, respectively.
    }
    \label{fig:floorplan_result}
\end{figure}

We illustrate the floorplan solutions in \Cref{fig:floorplan_result}. \Cref{fig:floorplan_result-a} and \Cref{fig:floorplan_result-c} show the results of a 16-chiplet floorplan and a 22-chiplet floorplan without performance consideration. 
The width of the line indicates the data movement frequency between chiplets, and the length of the line indicates the communication latency between chiplets. 
We omit the lines that represent low data movement frequency for clarity. 
\Cref{fig:floorplan_result-b} and \Cref{fig:floorplan_result-d} show the results of the second stage of floorplan optimization, which obviously reduces the inter-chiplet communication cost.
By reducing the HPWL wirelength by 0.57\% and the communication cost between chiplets by 24.81\%.

The performance-aware floorplan increases the overall area of the package by 0.86\%. This is the trade-off between the area cost and the performance improvement. The warpage also increases by 0.93\% compared to the floorplan without performance. However, these overheads are acceptable considering the significant communication cost reduction and obvious performance enhancement by 13.18\%.

Our framework improves the reliability of chiplet-based architecture by considering the warpage and bump stress issues in floorplan design and minimizing their effects. 
We compare our framework with the methods that do not consider bump stress and warpage issues, and the result is shown in \Cref{tab:reliability}. 

In our framework, we consider all aspects of performance, cost, area, and reliability of chiplet-based architecture. 
We demonstrate that ignoring performance metrics in floorplan design will degrade performance. 
By incorporating performance factors into floorplan design early, we realize the co-optimization of architecture and technology.

\section{Discussion and Limitation}\label{sec:discus}
\subsection{Comparison with Traditional Floorplan Algorithms}\label{sec:discus-tradition}

Most floorplan works like \cite{liu2014chiplet-floorplan,osmolovskyi2018chiplet-placement} do not consider reliability issues in the 2.5D package, \textit{i.e.}, considering the warpage threshold and avoiding the overlap between chiplets and the hotspot bumps.
In our work, the constraint of the bump stress is expressed in \Cref{equ: bump constraints} and shown in the left lower corner of \Cref{fig:chiplet-simulation}. 
After introducing reliability in our framework, methods like the enumeration-based algorithm \cite{liu2014chiplet-floorplan}, the branch-and-bound algorithms\cite{osmolovskyi2018chiplet-placement}, and the other traditional algorithms like B*Tree, SP, and CBL cannot fulfill the problem property.
The main differences between our framework and some traditional algorithms lay in the complexity of the problem constraints.

Therefore, adopting the mathematical programming (MP) methods in our framework can solve problems with multiple complex constraints well.
The MP methods can take complex constraints like warpage threshold, bump reliability, interposer area, and the wirelength into consideration simultaneously during the iterative optimization.
Meanwhile, the solutions to the MP problems can be finished by some excellent solvers like Gurobi\cite{2018gurobi}.
Therefore, we choose to compare our methods with MP-based methods in \cite{zhuang2022multi-package}, and the result illustrates that our methods outperform previous work and achieve a better performance-aware floorplan solution.

The timing complexity can not be analyzed precisely in the MP-solver-based optimization framework.
That is because the optimization problems in our framework are solved by existing the MP solver Gurobi, so the timing complexity is highly reliant on the scale of the chiplets in the problem.
However, we can obtain the running time of the solver for different tasks.
For simple chiplet-based architecture like $\mathcal{C}_{8},\mathcal{C}_{16}$, the whole optimization process can be finished in 20 minutes to 1 hour. 
However, for larger architecture like $\mathcal{C}_{22},\mathcal{C}_{30}$, to obtain a relative optimal solution will take 8-10 hours.
For some common chiplet-based architecture with 10 $\sim$ 30 chiplets, our framework can give a good solution in an acceptable time compared to some traditional methods like the simulated annealing algorithm.

\subsection{Limitation of the Framework}\label{sec:discus-thermal}

The thermal issue is important in 3D IC or chiplet-based 2.5D IC.
The main reason is the thermal can not be dissipated well with materials in the vertical direction.
Some works have built the thermal model in their work to simulate the heat flow in 3D IC.
An accurate thermal model is constructed based on the actual physical parameters of the packaging.
Our method focuses on introducing performance metrics into the floorplan stage to obtain an early optimization result, so the information related to thermal (\textit{e.g.}, voltage, accurate power) is not available in the current situation platform.

However, we plan to continue to develop the framework.
For future work, we will consider the thermal with a two-stage method in \cite{Chiplet-GIA-ICCAD2022-Li}.
Firstly, we will utilize a thermal simulation model to set the threshold, and after the first stage optimization, the \textbf{optChiplet} will continue to optimize the floorplan.
Even if our current framework doesn't take into account the thermal effect, we hope the idea of combining reliability, technology, and performance can bring more architecture insights into the IC manufacturing community.

\section{Conclusion}\label{sec:conc}

In this paper, we have presented Floorplet, a performance-aware framework for co-optimizing the floorplan and performance of chiplet-based architecture. 
We have addressed the challenges and drawbacks of using chiplets for complex circuit systems, such as degraded performance due to inter-chiplet communication, reliability issues due to warpage and bump stress, and lack of realistic chiplet designs for analysis.
We developed \textbf{parChiplet} to partition realistic SoC into functional chiplets, a simulation platform \textbf{simChiplet} to evaluate the performance impact of different floorplan solutions, and a floorplan framework \textbf{optChiplet} for chiplet-based architecture to consider performance, reliability, cost, and area metrics. 
We have tested our framework on commonly used benchmarks and shown its superiority over previous methods. 
Our work demonstrates the potential of using chiplets for designing high-performance and low-cost circuit systems. We believe that our framework can provide useful insights and guidance for future research and development of chiplet-based architecture.

{
    \bibliographystyle{IEEEtran}
    \bibliography{ref/Top,ref/reference}
}



\end{document}